\documentclass{elsart}

\usepackage{amssymb}
\usepackage{epsfig}
\usepackage{amsmath}
\usepackage{amsfonts}

\begin{document}

\begin{frontmatter}

\title{Numerical modeling of 1-D transient poroelastic waves in the low-frequency range}

\author{Guillaume Chiavassa$^a$}, 
\ead{guillaume.chiavassa@ec-marseille.fr}
\author{Bruno Lombard$^b$},
\ead{lombard@lma.cnrs-mrs.fr$^b$}
\ead[url]{http://w3lma.cnrs-mrs.fr/$\sim$MI/}
\author{Jo\"el Piraux$^b$}
\ead{piraux@lma.cnrs-mrs.fr}
\address{$^a$Ecole Centrale de Marseille and MSNM-GP, 13451 Marseille, France}
\address{$^b$Laboratoire de M\'ecanique et d'Acoustique, 13402 Marseille, France}

\begin{abstract}
Propagation of transient mechanical waves in porous media is numerically investigated in 1D. The framework is the linear Biot's model with frequency-independant coefficients. The coexistence of a propagating fast wave and a diffusive slow wave makes numerical modeling tricky. A method combining three numerical tools is proposed: a fourth-order ADER scheme with time-splitting to deal with the time-marching, a space-time mesh refinement to account for the small-scale evolution of the slow wave, and an interface method to enforce the jump conditions at interfaces. Comparisons with analytical solutions confirm the validity of this approach. 
\end{abstract}


\end{frontmatter}

\section{Introduction}\label{SecIntro}

The propagation of mechanical waves in porous media is of interest in many areas in applied mechanics, including industrial foams, spongious bones and petroleum rocks. The most-widely-used model describing the evolution of small mechanical perturbations in a saturated porous medium is that proposed by Biot in 1956. Two regimes were distinguished by Biot: one corresponding to a low-frequency range  \cite{BIOT56-A}, and one to a high-frequency range, where some of the physical parameters depend on the frequency \cite{BIOT56-B}. We focus on transient mechanical waves whose frequency content lies in the low-frequency range.

Until the mid 90's, Biot's equations were mainly studied in the harmonic regime. Various time-domain methods have been proposed since, based on finite-differences \cite{DAI95,ZENG01}, finite-elements \cite{THESE_EZZIANI,ZHAO05}, boundary-elements \cite{SCHANZ01}, and spectral methods \cite{CARCIONE99}. Since non-realistic values of the physical parameters were used, the real difficulties arising when performing time-domain simulations were often overlooked \cite{CARCIONE96,GUREVICH96}. These difficulties are induced by the coexistence of two solutions with radically different dynamics: a propagating "fast wave" and a diffusive "slow wave" \cite{CHANDLER81}. The latter is highly dispersed and attenuated, and remains localized near sources and interfaces. 

The aim of the present study is to develop an efficient numerical method to compute the solution made up of these two waves.  A time-splitting is used together with a fourth-order ADER scheme \cite{LORCHER05}. A flux-conserving space-time mesh refinement is implemented at the places where the diffusive wave is localized \cite{BERGER98}. Lastly, an immersed interface method gives a subcell resolution of the interfaces and accurately enforces the jump conditions between various materials \cite{PIRAUX01}. The numerical tools used are described and tested in 1D.

The paper is organized as follows. The Biot's model is briefly recalled in section 2. The numerical tools are described in section 3. Section 4 presents numerical experiments confirming the validity and efficiency of this approach. In section 5, conclusions are drawn and some future perspectives are suggested.

\section{Problem statement}\label{SecState}

\subsection{Biot's model in the low-frequency range}\label{SubSecBiot}

Biot's model describes the propagation of mechanical waves in a porous medium consisting of a solid matrix saturated with fluids circulating freely through the pores. The underlying hypotheses in the low-frequency range are as follows:
\begin{itemize}
\item the wavelength of the perturbations is large in comparison with the diameter of the pores, as well as with the representative macroscopic volumes;
\item the amplitudes of the perturbations are small;
\item the elastic and isotropic matrix is fully saturated by a single fluid phase.
\end{itemize}
The model is based on 10 physical parameters: the density $\rho_f$ and the dynamic viscosity $\eta$ of the fluid; the density $\rho_s$ and the shear modulus $\mu$ of the solid; the porosity $0<\phi<1$, the tortuosity $a\geq 1$, the absolute permeability $\kappa$, the Lam\'e coefficient $\lambda_f$, and the two Biot coefficients $\beta$ and $m$ of the saturated matrix. The following notation is introduced: $\rho_w=\frac{a}{\phi}\,\rho_f$, $\rho=\phi\,\rho_f+(1-\phi)\,\rho_s$, and $\chi =\rho\,\rho_w-\rho_f^2>0$. The unknowns are the elastic velocity $v_s$, the elastic stress $\sigma$, the filtration velocity $w=\phi\,(v_f-v_s)$, where $v_f$ is the fluid velocity, and the acoustic pressure $p$. When dealing with heterogeneous media, all of the parameters are assumed to be piecewise constant and discontinuous across the interfaces. Denoting $x=\alpha$ the location of the interface and $\kappa_s$ its hydraulic permeability, the jump conditions are \cite{GUREVICH99}:
\begin{equation}
[v_s(\alpha,\,t)]=0,\quad [w(\alpha,\,t)]=0,\quad [\sigma(\alpha,\,t)]=0,\quad [p(\alpha,\,t)]=-\frac{\textstyle 1}{\textstyle \kappa_s }\,w(\alpha^-,\,t).
\label{JCscal}
\end{equation} 
Since $w$ is continuous, no privileged orientation is induced by the last condition in (\ref{JCscal}). If $\kappa_s\rightarrow+\infty$, the open-pore conditions are obtained, with perfect hydraulic contact \cite{BOURBIE87}. If  $\kappa_s\rightarrow 0$, no motion of the fluid relative to the matrix occurs: the pores are closed.  

In the case of a harmonic wave of frequency $f$, the low-frequency Biot's theory is valid as long as $f<0.15\,f_c$ \cite{BIOT56-B}, with
\begin{equation}
f_c=\frac{\textstyle \eta\,\phi}{\textstyle 2\,\pi\,a\,\kappa\,\rho_f}.
\label{Fc}
\end{equation}
When $f\geq 0.15 \,f_c$, the width of the viscous boundary layer is smaller than that of pores, and the flow of fluid is no more of Poiseuille's type. Consequently, a non-polynomial frequential correction is required in the dynamic permeability $\kappa/\eta$, leading to fractional derivatives in the time-domain \cite{HANYGA05}.

\subsection{Initial boundary-value problem}\label{IBVP}

Setting $\boldsymbol{U}=(v_s,\,w,\,\sigma,\,p)^T$, 
\begin{equation}
\boldsymbol{A}=
\left(
\begin{array}{cccc}
0 & 0 & -\rho_w/\chi & -\rho_f/\chi \\
0 & 0 & \rho_f/\chi & \rho/\chi \\
-(\lambda_f+2\,\mu) & -\beta\,m & 0 & 0\\
\beta\,m & m & 0 & 0
\end{array}
\right),\quad
\boldsymbol{S}=
\frac{\textstyle \eta}{\textstyle \kappa}\frac{\textstyle \rho}{\textstyle \chi } 
\left(
\begin{array}{cccc}
0 & -\rho_f/\rho & 0 & \hspace{0.4cm} 0\\ 
0 & 1 & 0 & \hspace{0.4cm} 0\\ 
0 & 0 & 0 & \hspace{0.4cm} 0\\
0 & 0 & 0 & \hspace{0.4cm} 0
\end{array}
\right),
\label{MatA}
\end{equation}
the Biot's evolution equations \cite{BIOT56-A} are written together with the jump conditions (\ref{JCscal}) in the form of a first-order linear system with a source term
\begin{equation}
\left\{
\begin{array}{l}
\displaystyle
\frac{\textstyle \partial}{\textstyle \partial\,t}\,\boldsymbol{U}+\frac{\textstyle \partial}{\textstyle \partial\,x}\,\boldsymbol{A}\, \boldsymbol{U}=-\boldsymbol{S}\,\boldsymbol{U}\quad \mbox{if }\, x\neq\alpha, t\geq 0,\\
[10pt]
\displaystyle
[\boldsymbol{C}\,\boldsymbol{U}(\alpha,\,t)]=\boldsymbol{0},\\
[10pt]
\displaystyle
\boldsymbol{U}(x,\,0)=\boldsymbol{U}_0(x),
\end{array}
\right.
\label{LC}
\end{equation}
where the $4\times 4$ matrices $\boldsymbol{C}$ are derived from (\ref{JCscal}). A detailed mathematical analysis of the solution to (\ref{LC}) in a homogeneous medium can be found in \cite{THESE_EZZIANI}. Here we restrict the discussion to basic properties that are directly useful for numerical modeling purposes. The spectral radius of $\boldsymbol{S}$ is $R(\boldsymbol{S})=\frac{\eta}{\kappa}\,\frac{\rho}{\chi}$. The eigenvalues of $\boldsymbol{A}$ are real and denoted $\pm \overline{c}_1$ and $\pm \overline{c}_2$, with $\overline{c}_1 > \overline{c}_2 >0$. Injecting a mode $e^{i(\omega\,t-k\,x)}$ in (\ref{LC}), where $\omega$ is the angular frequency and $k$ the wavenumber, gives the dispersion relation
\begin{equation}
\begin{array}{l}
\displaystyle
A\,k^4+B(\omega)\,k^2+C(\omega)=0,\\
[8pt]
\displaystyle
A=\kappa\,m\left(\lambda_f+2\,\mu-\beta^2\,m\right),\\
[8pt]
\displaystyle
B(\omega)=-\kappa\left(\left(\lambda_f+2\,\mu\right)\,\rho_w+m\left(\rho-2\,\beta\,\rho_f\right)\right)\omega^2+i\,\eta\left(\lambda_f+2\,\mu\right)\omega,\\
[8pt]
\displaystyle
C(\omega)=\chi\kappa\,\omega^4-i\,\eta\,\rho\,\omega^3,
\end{array}
\label{Dispersion}
\end{equation}
with roots $\pm k_1$ and $\pm k_2$, $\Re\mbox{e}\left\{k_{1,2}\right\}>0$. The phase velocities $c_1=\omega\,/\,\Re\mbox{e}\left\{k_1\right\}$ and $c_2=\omega\,/\,\Re\mbox{e}\left\{k_2\right\}$ are strictly increasing functions of frequency, tending in the high-frequency limit towards $\overline{c}_1$ and $\overline{c}_2$. The corresponding waves are called the fast wave and the slow wave, respectively. If $\eta=0$, which is physically irrelevant with usual media, the waves are purely propagated and the mechanical energy is constant. If $\eta \neq 0$, the fast wave is almost non-dispersive and non-diffusive. On the contrary, the slow wave becomes highly dispersive and diffusive. If $f\ll f_c$, then $c_2\ll \overline{c}_2$: the slow wave tends towards a non-propagating mode \cite{CHANDLER81}. The direct contribution of the slow wave to the overall wave propagation processes is therefore negligible. However, the accurate computation of the fast wave depends crucially on the effects of the slow wave on the balance equations at interfaces \cite{BOURBIE87}. 

\section{Numerical tools}\label{SecNum}

\subsection{Numerical scheme}\label{SubSecSchem}

This subsection deals with the numerical resolution of (\ref{LC}) far from $\alpha$, that is where the stencil does not cross the interface. A uniform grid  is considered here, with the spatial mesh size $\Delta\,x$ and the time step $\Delta\,t$. An approximation $\boldsymbol{U}_i^n$ of $\boldsymbol{U}(x_i=i\,\Delta\,x,\,t_n=n\,\Delta\,t)$ is sought. The numerical methods recalled in section \ref{SecIntro} usually consist of solving simultaneously the propagating part and the source term in (\ref{LC}). If $\eta \neq 0$, a Von-Neumann analysis of stability typically yields
\begin{equation}
\Delta\,t\leq\min\left(\frac{\textstyle \Delta\,x}{\textstyle \overline{c}_1},\,\frac{\textstyle 2}{\textstyle  R(\boldsymbol{S})}\right),
\label{CFL}
\end{equation}
which is highly restrictive since $R(\boldsymbol{S})$ may be large. For instance, $\mbox{CFL}=\overline{c}_1\,\Delta\,t/\Delta\,x\approx 10^{-2}$ with the media considered in section (\ref{viscMed}), and $\rm{CFL} \approx 10^{-12}$ can be reached with highly dissipative fluids such as bitumen. A much more efficient approach is to split (\ref{LC}) and to alternatively solve by Strang's splitting \cite{LEVEQUE07} the propagating part
\begin{equation}
\frac{\textstyle \partial}{\textstyle \partial\,t}\,\boldsymbol{U}+\frac{\textstyle \partial}{\textstyle \partial\,x}\,\boldsymbol{A}\,\boldsymbol{U}=\boldsymbol{0},
\label{SplittingA}
\end{equation}
and the source term part
\begin{equation}
\frac{\textstyle \partial}{\textstyle \partial\,t}\,\boldsymbol{U}=-\boldsymbol{S}\,\boldsymbol{U}.
\label{SplittingB}
\end{equation}
The equation (\ref{SplittingA}) can be solved by applying any explicit two time step spatially-centered flux-conserving scheme. The latter can be written abstractly
\begin{equation}
\boldsymbol{U}_i^{n+1/2} = \boldsymbol{H}\left(\boldsymbol{U}_{i-s}^n,...,\,\boldsymbol{U}_{i+s}^n\right),
\label{TM}
\end{equation}
where $s$ is the width of the stencil. In the present numerical experiments, a temporally and spatially  fourth-order accurate ADER scheme was used \cite{LORCHER05,SCHWARTZKOPFF04}, with $s=2$. This scheme accounts for waves over long distances with small dispersion and diffusion errors, even on coarse grids. The equation (\ref{SplittingB}) is solved exactly: $p$ and $\sigma$ are unchanged, whereas the velocities become 
\begin{equation}
v_i^{n+1}=v_i^{n+1/2}+\frac{\rho_f}{\rho}\left(1-e^{-\frac{\eta}{\kappa}\,\frac{\rho}{\chi}\,\frac{\Delta\,t}{2}}\right)w_i^{n+1/2},\qquad w^{n+1}_i=e^{-\frac{\eta}{\kappa}\,\frac{\rho}{\chi}\,\frac{\Delta\,t}{2}}\,w_i^{n+1/2}. 
\label{Filtration}
\end{equation}
The splitting (\ref{SplittingA})-(\ref{SplittingB}) together with the exact integration (\ref{Filtration}) yields the optimal stability condition $\mbox{CFL}\leq 1$. However, since $\boldsymbol{A}$ and $\boldsymbol{S}$ do not commute, it decreases the theoretical order of convergence from 4 to 2. In practice, the convergence rate measured was close to 3 in our experiments.

\subsection{Mesh refinement}\label{SubSecMesh}

If $\eta \neq 0$ and $f\ll f_c$, the slow wave has much smaller spatial scales of evolution than the wavelength of the fast wave. A very fine grid is consequently required to account for its evolution. Since the use of a fine uniform grid on the whole computational domain is out of reach in view of 2-D simulations, the grid refinement procedure provides a good alternative. In addition, since the slow wave remains localized near the sources and the interfaces, grid refinement is necessary only around these places.
 
To limit the numerical dispersion on the coarse grid, it is preferable to also perform temporal refinement using a local CFL stability condition. Many algorithms for space-time mesh refinement have been recently developed in a variational framework, which ensure the stability of the coupling between grids by conserving a discrete energy \cite{RODRIGUEZ05}. Here we adopt another approach based on flux conservation \cite{BERGER98}, which is more naturally coupled to the flux-conserving scheme (\ref{TM}). The extrapolated values required to couple coarse and fine grids are obtained by performing linear interpolation in space and time on the numerical values at the surrounding nodes. In the case of the Lax-Wendroff scheme applied to the advection equation, the stability of the coupling was proved in \cite{BERGER85} for arbitrary refinement factor.

The mesh refinement with a factor $q$ on a zone of width $h$ introduces a computational extra-cost proportional to the product of grid nodes $h\,q$ by the number of time substeps $q$. Consequently, one must estimate carefully $q$ and $h$ in terms of the physical parameters, in order to get efficient simulations. For that purpose, one uses the wavelengths $\lambda_1=c_1/f$ and $\lambda_2=c_2/f$ deduced from (\ref{Dispersion}). The numbers of grid nodes per wavelength of the slow wave, in the fine grid, and of the fast wave, on the coarse grid, must be roughly equal, which provides $q\approx c_1/c_2$. This factor may be large: in section \ref{SecExp}, one gets $q \approx 64$. Numerical experiments have shown that it is preferable to use intermediate smaller refinement factors, for instance 1 to 8 and then 8 to 64. Concerning $h$, the refined zone must contain a few wavelengths of the slow wave, leading typically to $h=4\,\lambda_2$ in section \ref{SecExp}.

\subsection{Interface method}\label{SubSecMI}

\begin{figure}[htbp]
\begin{center}
\begin{tabular}{c}
\includegraphics[scale=0.7]{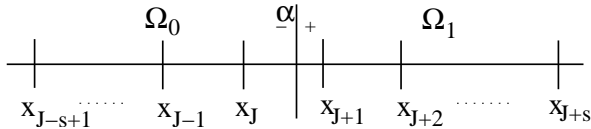}
\end{tabular}
\end{center}
\caption{Interface between media $\Omega_0$ and $\Omega_1$ and the irregular nodes around $\alpha$.} 
\label{FigInterface}
\end{figure}

This subsection deals with the time-marching at the grid nodes where the stencil of (\ref{TM}) crosses $\alpha$. We define $J$ by $x_J \leq \alpha <x_{J+1}$, and we assume that the grid is uniform from $x_{J-s+1}$ to $x_{J+s}$ (figure \ref{FigInterface}). At the so-called irregular nodes $x_{J-s+1},...,x_{J+s}$, the scheme (\ref{TM}) must not be applied na\"ively, for three reasons. First, the spatial derivatives of the solution are not smooth across $\alpha$, hence the order of convergence of the scheme decreases. Secondly, the subcell position of $\alpha$ inside the mesh is ignored, which leads to a $O(\Delta\,x)$ error. Thirdly, the jump conditions (\ref{JCscal}) are not enforced in the scheme. The numerical waves diffracted by the interface may therefore not tend towards the exact waves.

At the irregular nodes, we adapt an immersed interface method \cite{PIRAUX01}. This method requires knowing the jump conditions satisfied by the spatial derivatives of $\boldsymbol{U}$. We deduce these conditions from (\ref{LC}) for all $m\geq 1$
\begin{equation}
\begin{array}{lll}
\displaystyle \frac{\textstyle \partial^m}{\textstyle \partial\,t^m}\left[\boldsymbol{C}\,\boldsymbol{U}(\alpha,\,t)\right]&=&
\displaystyle
\left[(-1)^m\boldsymbol{C}\,
\left(
\boldsymbol{A}\,\frac{\textstyle \partial}{\textstyle \partial\,x}+\boldsymbol{S}
\right)^m
\boldsymbol{U}(\alpha,\,t)
\right]=\boldsymbol{0},\\
[8pt]
& \Rightarrow & 
\displaystyle
\left(\boldsymbol{U},...,\,\frac{\textstyle \partial^m}{\textstyle \partial\,x^m}\boldsymbol{U}\right)(\alpha^+,\,t)=\boldsymbol{D}_m\left(\boldsymbol{U},...,\frac{\textstyle \partial^m}{\textstyle \partial\,x^m}\boldsymbol{U}\right)(\alpha^-,\,t),
\end{array}
\label{JCm}
\end{equation}
where $\boldsymbol{D}_m$ is a $4\,(m+1)\times 4\,(m+1)$ matrix depending on the permeability $\kappa_s$ and on the physical parameters around $\alpha$. Let $k$ be a positive integer. At the $2\,k$ grid nodes surrounding $\alpha$, $2\,k$-th order Taylor expansions of $\boldsymbol{U}(x_i,\,t_n)$ on $\alpha^\pm$, together with the jump conditions (\ref{JCm}), are written in the matrix form
\begin{equation}
\left(
\begin{array}{c}
\boldsymbol{U}(x_{J-k+1},\,t_n)\\
\vdots\\
\displaystyle
\boldsymbol{U}(x_{J+k},\,t_n)
\end{array}
\right)
=\boldsymbol{M}
\left(
\begin{array}{c}
\boldsymbol{U}(\alpha^-,t_n)\\
\vdots\\
\displaystyle
\frac{\textstyle \partial ^{2\,k-1}}{\textstyle \partial \,x ^{2\,k-1}}\,\boldsymbol{U}(\alpha^-,\,t_n)
\end{array}
\right)
+
\left(
\begin{array}{c}
\boldsymbol{O}(\Delta \,x^{2\,k})
\\
\vdots\\
\displaystyle
\boldsymbol{O}(\Delta \,x^{2\,k})
\end{array}
\right),
\label{Taylor1}
\end{equation}
where $\boldsymbol{M}$ is a $8\,k\times 8\,k$ matrix. The Taylor series remainder term is removed from (\ref{Taylor1}), and the exact values are replaced by numerical ones. The spatial derivatives estimated by performing inversion of (\ref{Taylor1}) are used to build smooth extensions of the solution, called modified values, on the right of $\alpha$:
\begin{equation}
i=J+1,...,J+s,\qquad
\boldsymbol{U}_i^*=\left(\boldsymbol{I}_4,...,\frac{\textstyle(x_i-\alpha)^{2\,k-1}}{\textstyle (2\,k-1)\,!} \boldsymbol{I}_4\right)\boldsymbol{M}^{-1}
\left(
\begin{array}{c}
\boldsymbol{U}_{J-k+1}^n\\
\vdots\\
\displaystyle
\boldsymbol{U}_{J+k}^n
\end{array}
\right),
\label{SolMod}
\end{equation}
where $\boldsymbol{I}_4$ is the $4\times 4$ identity matrix. Modified values on the left of $\alpha$ are defined similarly. They are then injected into the scheme at the irregular nodes:
\begin{equation}
\begin{array}{l}
i=J-s+1,...,J,\quad 
\boldsymbol{U}_i^{n+1/2}=\boldsymbol{H}_{\Omega_0}\left(\boldsymbol{U}_{i-s}^n,...,\,\boldsymbol{U}_J^n,\,\boldsymbol{U}_{J+1}^*,...,\,\boldsymbol{U}_{i+s}^*\right),\\
[8pt]
i=J+1,...,J+s,\quad
\boldsymbol{U}_i^{n+1/2}=\boldsymbol{H}_{\Omega_1}\left(\boldsymbol{U}_{i-s}^*,...,\,\boldsymbol{U}_J^*,\,\boldsymbol{U}_{J+1}^n,...,\,\boldsymbol{U}_{i+s}^n\right),
\end{array}
\label{TMESIM}
\end{equation}
where $\boldsymbol{H}_{\Omega_0}$ and $\boldsymbol{H}_{\Omega_1}$ denote the operator $\boldsymbol{H}$ with physical parameters of media $\Omega_0$ and $\Omega_1$, respectively. Some comments about the interface method:
\begin{itemize}
\item since the jump conditions are linear, the work is mainly carried out during a preprocessing step. At each time step, only small matrix-vector products (\ref{SolMod}) are required to compute the modified values. The computational cost is therefore negligible in comparison with that of time-marching;
\item the matrix $\boldsymbol{M}$ in (\ref{Taylor1}) depends on the jump conditions and on the position of $\alpha$ inside the mesh. Using the modified values (\ref{TMESIM}) introduces into the scheme a subcell resolution of the interface, which removes the $O(\Delta\,x)$ error \cite{PIRAUX01};
\item in the limit case where the parameters are the same on both sides of $\alpha$ and the hydraulic contact is perfect, $\boldsymbol{U}_i^*=\boldsymbol{U}_i^n$ if $k\geq s$. Consequently, we again obtain the scheme applied in homogeneous medium;
\item with a $r$-th order accurate scheme, the local truncation error of (\ref{TMESIM}) at irregular nodes is $r$-th order if $2\,k-1\geq r$ \cite{PIRAUX01}. The fourth-order ADER scheme therefore requires $k=3$. As deduced from \cite{GUSTAFSSON75}, $k=2$ suffices to ensure fourth-order overall accuracy;
\item GKS analysis \cite{GKS72} has been performed in the case of inviscid fluids on a uniform grid to determine the stability of the hybrid scheme (\ref{TM}) and (\ref{TMESIM}). This analysis is based on the possible existence of discrete increasing modes emitted solely by the interface without any incident field \cite{TREFETHEN83}. A parametric study showed that the hybrid scheme is stable in tests 1 and 2 presented below, whatever the position of the interface, with $k=1,2,3$. 
\end{itemize}

\section{Numerical experiments}\label{SecExp}

\begin{table}[htbp]
\begin{center}
\begin{tabular}{l|ll|ll}
Parameters & $\Omega_0$ & $\Omega_1$ & $\Omega_0$ & $\Omega_1$\\
\hline
$\rho_f$ (kg/m$^3$)    & 1040         & 1040            & 1040         & 10             \\ 
$\eta$ (Pa.s)          & 0            & 0               & $10^{-3}$    & $2.2\,10^{-5}$ \\
$\rho_s$ (kg/m$^3$)    & 2650         & 2211            & 2650         & 2650           \\
$\mu$ (Pa)             & $1.85\,10^9$ & $3.54\,10^9$    & $1.85\,10^9$ & $1.85\,10^9$   \\
$\phi$                 & 0.3          & 0.01            & 0.3          & 0.3            \\
$a$                    & 2            & 2               & 2            & 2              \\
$\kappa$ (m$^2$)       & $10^{-12}$   & $10^{-16}$      & $10^{-12}$   & $10^{-12}$     \\
$\lambda_f$ (Pa)       & $8.40\,10^9$ & $4.69\,10^9$    & $8.40\,10^9$ & $2.43\,10^9$   \\
$\beta$                & 0.88         & 0.01            & 0.88         & 0.35           \\
$m$ (Pa)               & $7.05\,10^9$ & $2.46\,10^{11}$ & $7.05\,10^9$ & $5.37\,10^7$   \\
\hline
$\overline{c}_1$ (m/s) & 2364.9       & 2314.1          & 2364.9       & 1817.9         \\
$\overline{c}_2$ (m/s) & 774.9        & 1087.7          & 774.9        & 897.6          \\
$c_1$ (m/s)            & 2364.9       & 2314.1          & 2364.5       & 1817.8         \\
$c_2$ (m/s)            & 774.9        & 1087.7          & 38.1         & 30.3           \\
$c_1\,/\,c_2$          & 3.05         & 2.12            & 62.06        & 60.0           \\
$f_c$ (Hz)             & 0            & 0               & 22955        & 52521
 
\end{tabular}
\caption{Parameters and related data in tests 1-2 (central column) and 3-6 (right column).}
\label{TabParametres}
\end{center}
\end{table}

A 400-m domain is studied. Incident, reflected and transmitted waves are denoted by I, R and T. Fast and slow waves are denoted by F and S. Except in test 6, analytical and numerical solutions are shown in solid lines and circles. Vertical dotted lines denote the position of mesh refinement. If $\eta=0$ in each media, computing the analytical solution is straightforward. Otherwise, it follows from a Fourier analysis. The source is a $C^4$ truncated sinusoid with a central frequency $f=30$ Hz, as shown in Figure \ref{Test1}-a. On the main grid, $\Delta\,x=1$ m, and the computations are performed with $\mbox{CFL}=0.9$.

Two sets of physical parameters are used, shown in table \ref{TabParametres}. In tests 1 and 2, they model sandstone saturated with water ($\Omega_0$) and schist saturated with water ($\Omega_1$), except that $\eta=0$ \cite{DAI95}. This is not physically realistic, but it sheds light on the ADER scheme and on the interface method: $\boldsymbol{S}=\boldsymbol{0}$ means that no splitting (\ref{SplittingB}) is required. Nor is mesh refinement required, since the slow wave propagates. In tests 3 to 6, the parameters model sandstone saturated with water ($\Omega_0$) and with gas ($\Omega_1$). Since $\eta\neq 0$ and $f \ll f_c$, the slow wave is a static mode, which puts the focus on the mesh refinement. The refinement factor deduced from $c_1/c_2$ is 64: see section \ref{SubSecMesh} and table \ref{TabParametres}. 

\subsection{Inviscid media}

\begin{figure}[htbp]
\begin{center}
\begin{tabular}{cc}
(a) & (b)\\
\includegraphics[scale=0.32]{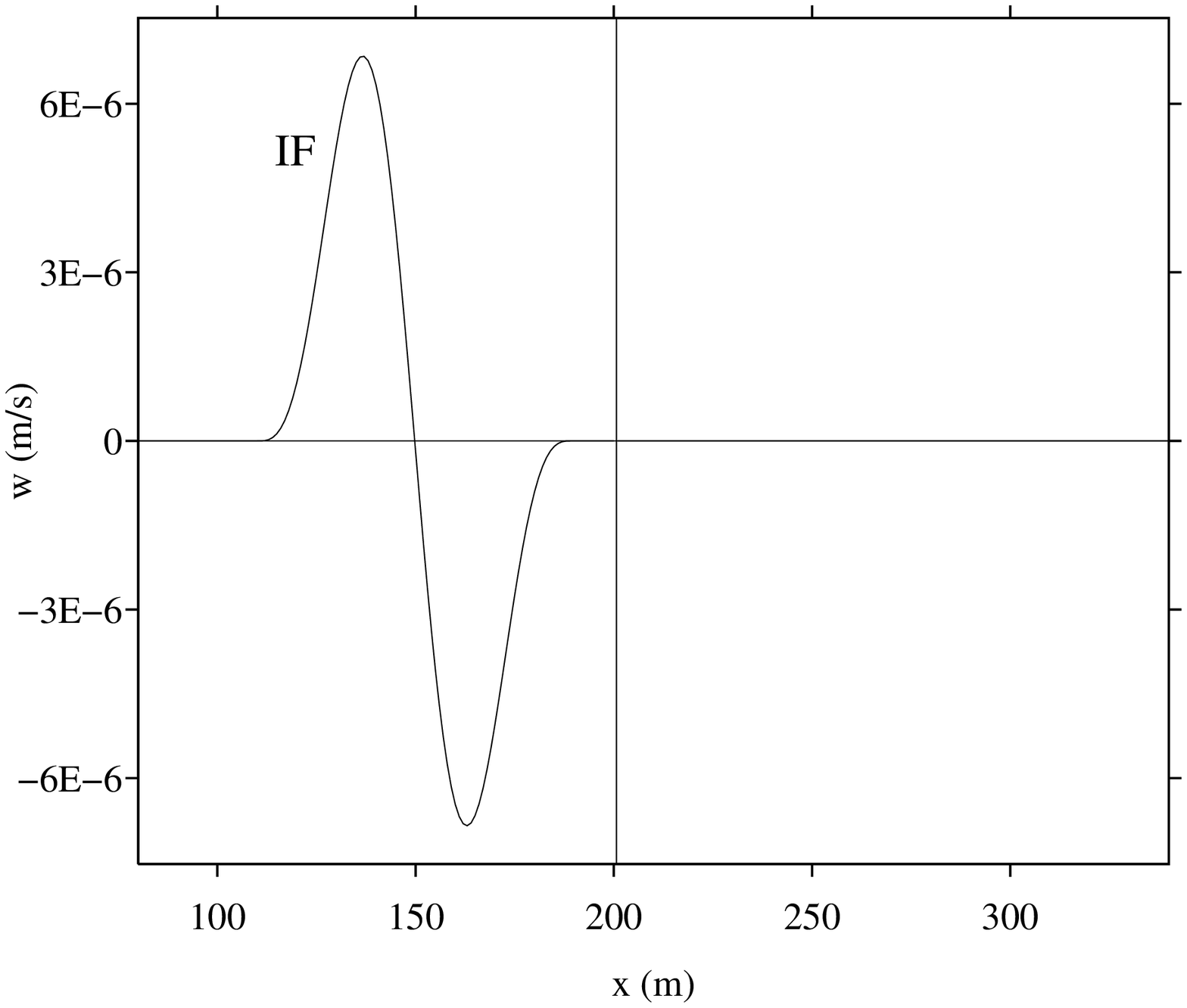}&
\includegraphics[scale=0.32]{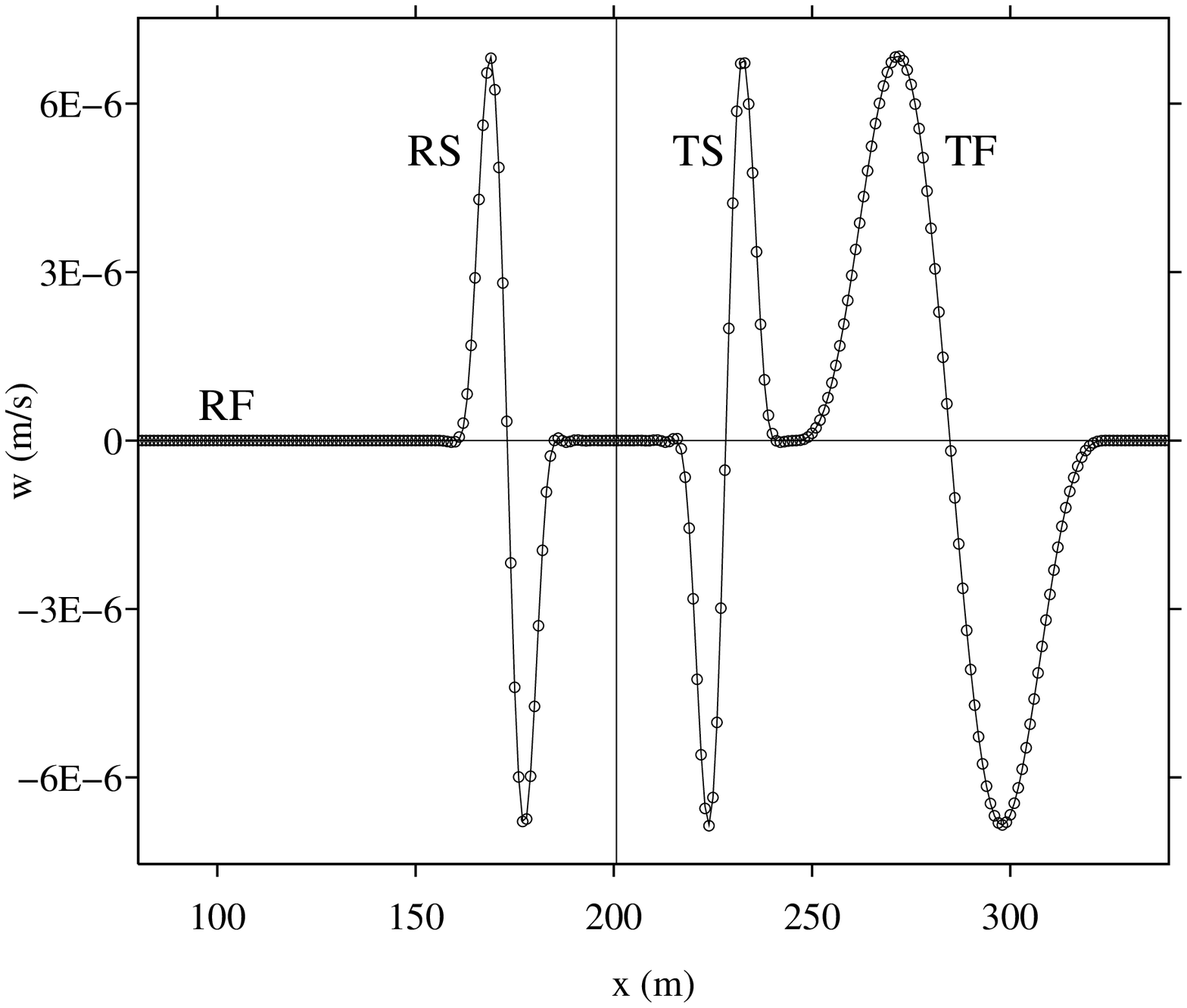}
\end{tabular}
\end{center}
\caption{Test 1: at initial instant (a) and after crossing the interface (b).} 
\label{Test1}
\end{figure}

First we consider identical media $\Omega_0$ linked by an imperfect hydraulic contact $\kappa_s=10^{-16}$ m.s$^{-1}$.Pa$^{-1}$ at $\alpha=200.67$ m. Wave conversions are shown in Figure \ref{Test1}-b, and a good agreement is seen between the analytical and numerical values; on this scale, the reflected fast wave is not visible. Since no contrast of Biot's coefficients occurs, the unmodified scheme (\ref{TM}) without the interface method (\ref{TMESIM}) would propagate the incident wave without diffraction. 

\begin{figure}[htbp]
\begin{center}
\begin{tabular}{cc}
(a) & (b)\\
\includegraphics[scale=0.32]{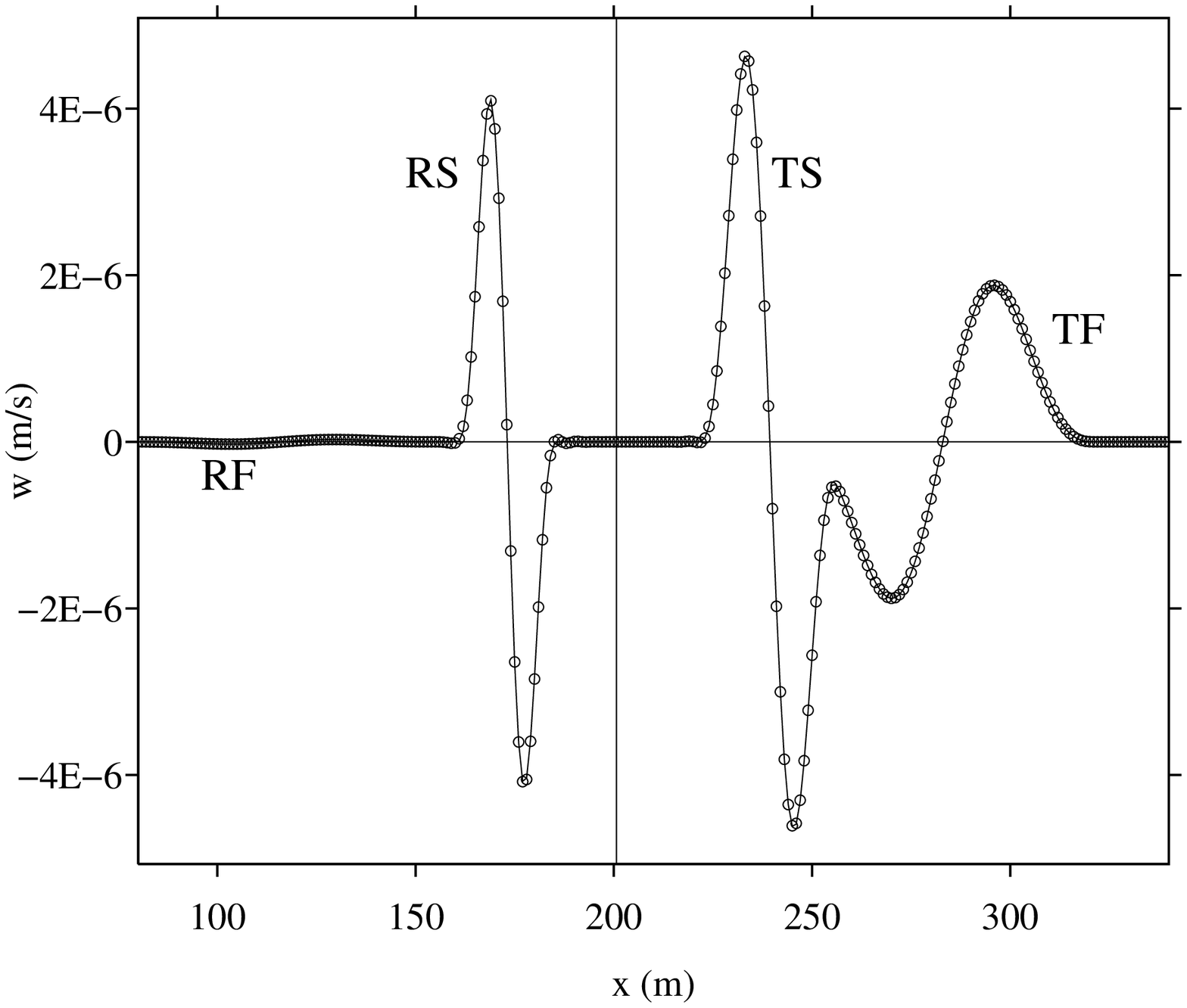}&
\includegraphics[scale=0.32]{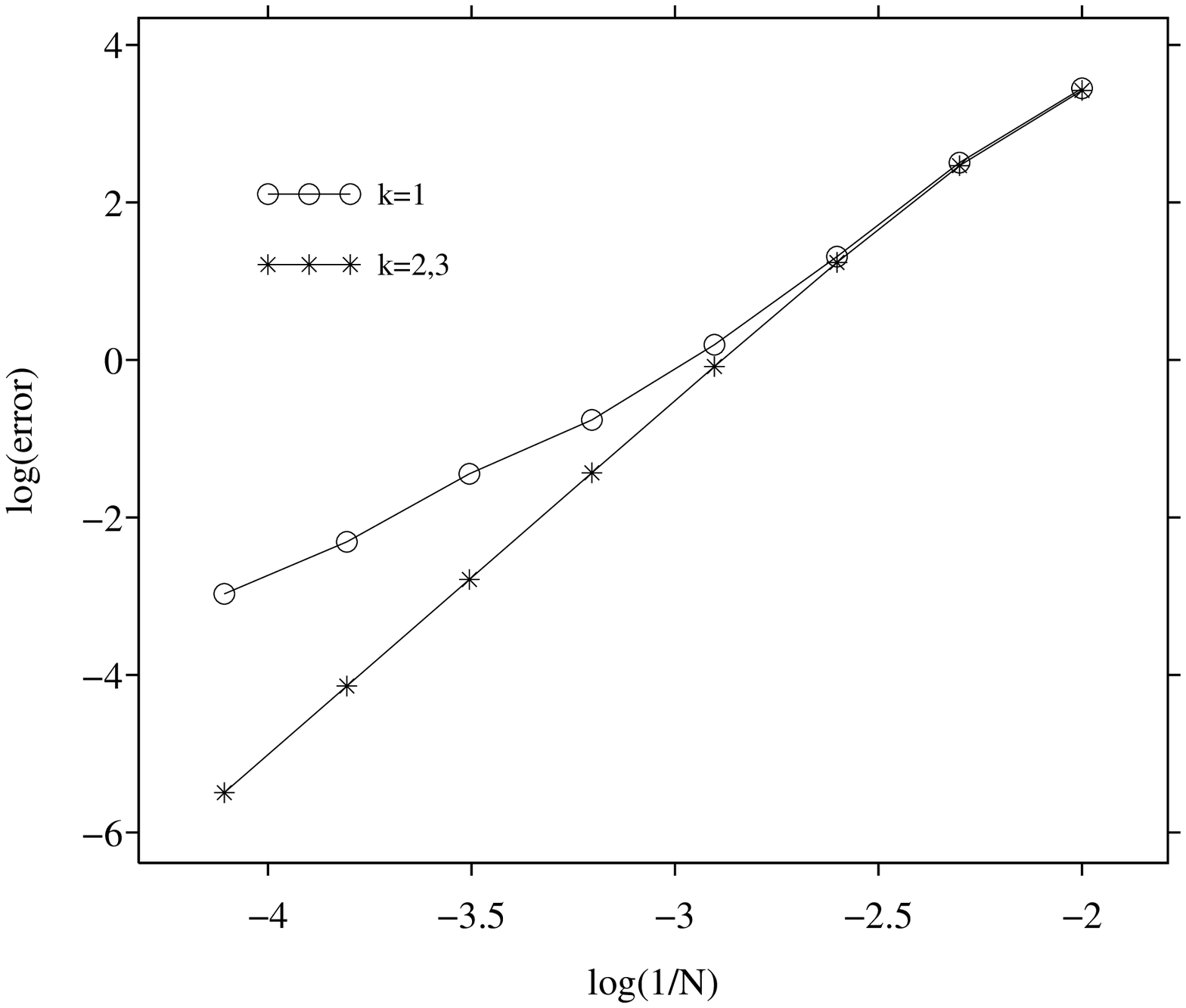}
\end{tabular}
\end{center}
\caption{Test 2: after crossing the interface (a) and convergence measurements (b).} 
\label{Test2}
\end{figure}

Test 2 deals with media ($\Omega_0,\Omega_1$) linked by a perfect hydraulic contact $1/\kappa_s=0$. Without using the interface method, GKS analysis and direct simulations show that the scheme is unstable and the solution grows exponentially when crossing $x=\alpha$. Figure \ref{Test2}-a gives the analytical and numerical values obtained. The results of convergence studies are presented in Figure \ref{Test2}-b. As mentioned in section \ref{SubSecMI}, $k=2$ or $k=3$ maintain the fourth-order accuracy of the ADER scheme; $k=1$ does not suffice, leading to a rate of convergence of only 2.9.

\begin{figure}[htbp]
\begin{center}
\begin{tabular}{cc}
(a) & (b)\\
\includegraphics[scale=0.32]{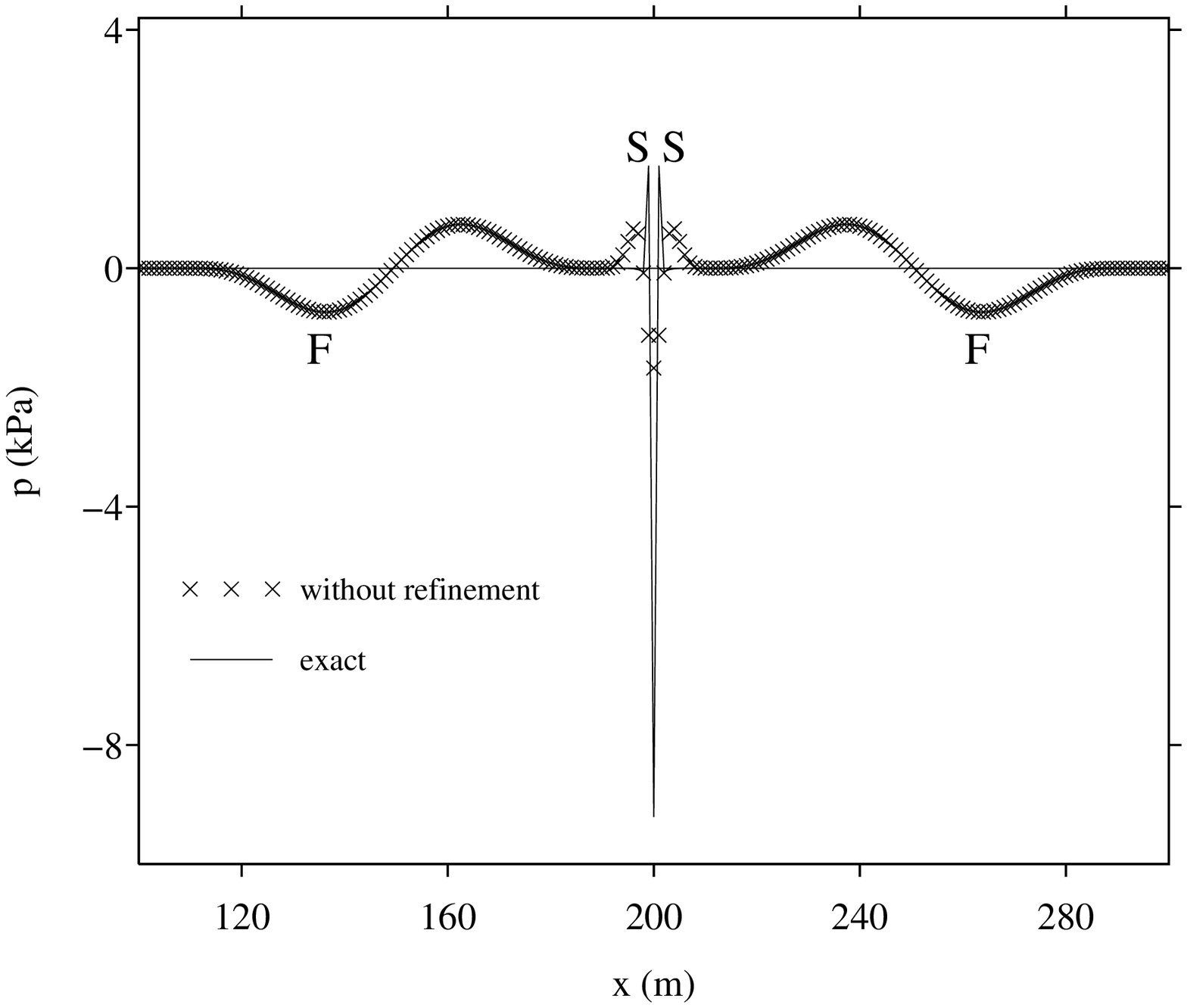}&
\includegraphics[scale=0.32]{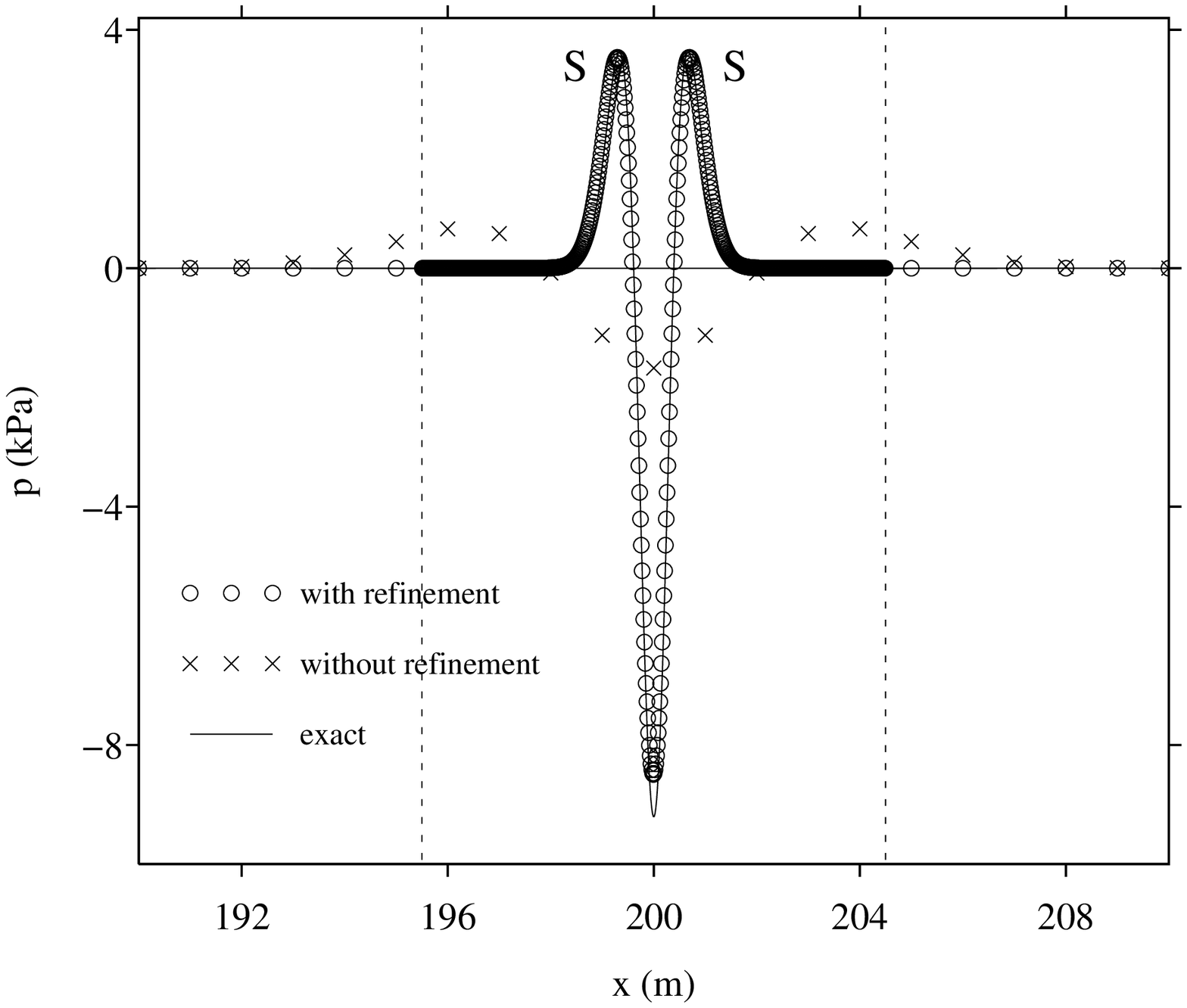}
\end{tabular}
\end{center}
\caption{Test 3: no refinement (a); zoom around $x_s$, with and without refinement (b).} 
\label{Test3}
\end{figure}

\subsection{Dissipative media} \label{viscMed}

Test 3 focuses on the homogeneous dissipative medium $\Omega_0$ excited by a ponctual stress source of finite duration at $x_s=200$ m. A direct discretization with no splitting would give $\mbox{CFL}=0.03$ (\ref{CFL}). A snapshot of $p$ is shown in figure \ref{Test3}-a. Fast waves are advected rightwards and leftwards while the slow waves remain localized around $x_s$, and vary considerably on small spatial scales. Their numerical values are highly smeared. In figure \ref{Test3}-b, the coarse grid solution is compared with a solution refined 64 times around $x_s$. Good agreement is seen between the exact and refined numerical values.

\begin{figure}[htbp]
\begin{center}
\begin{tabular}{cc}
(a) & (b)\\
\includegraphics[scale=0.32]{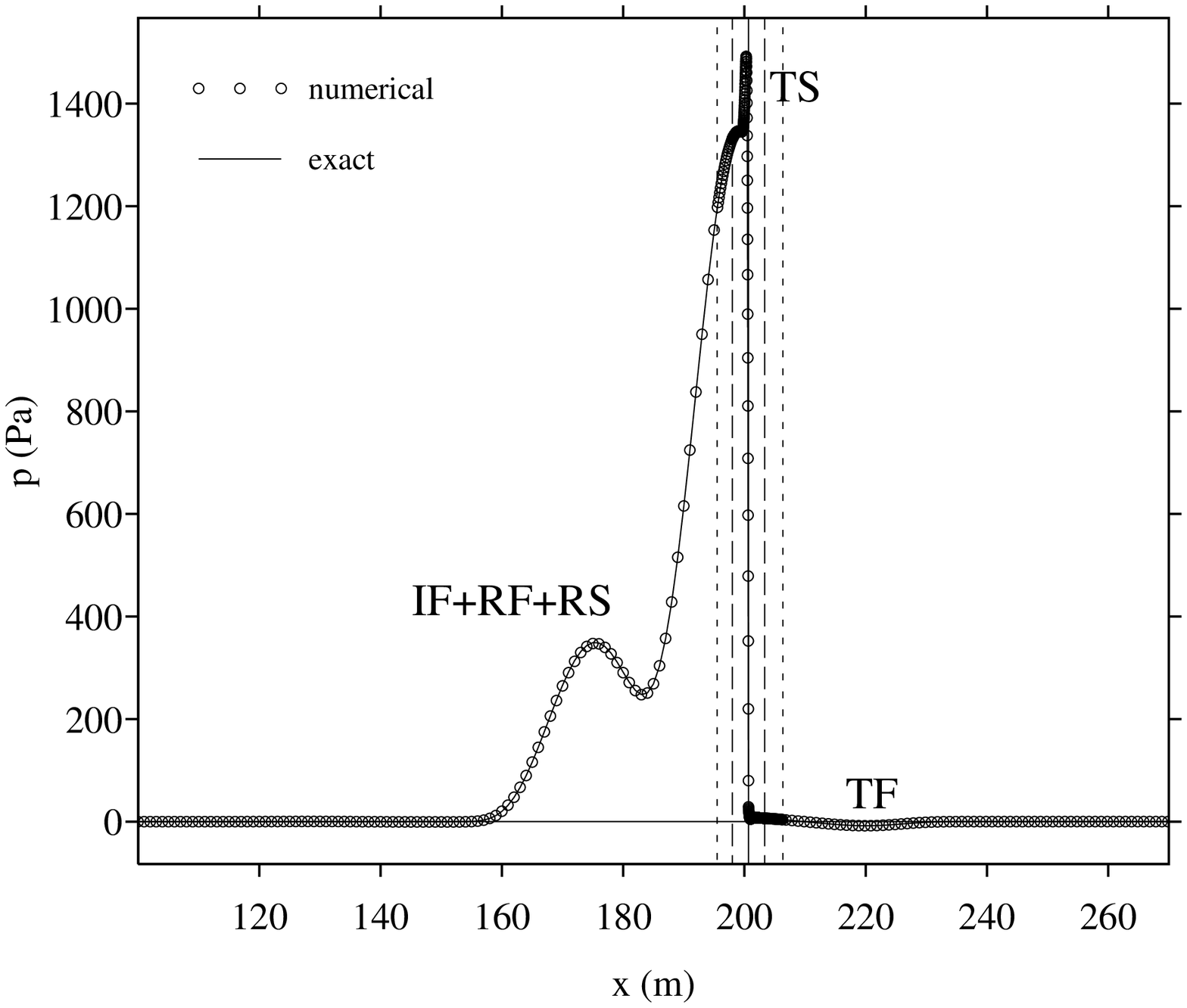}&
\includegraphics[scale=0.32]{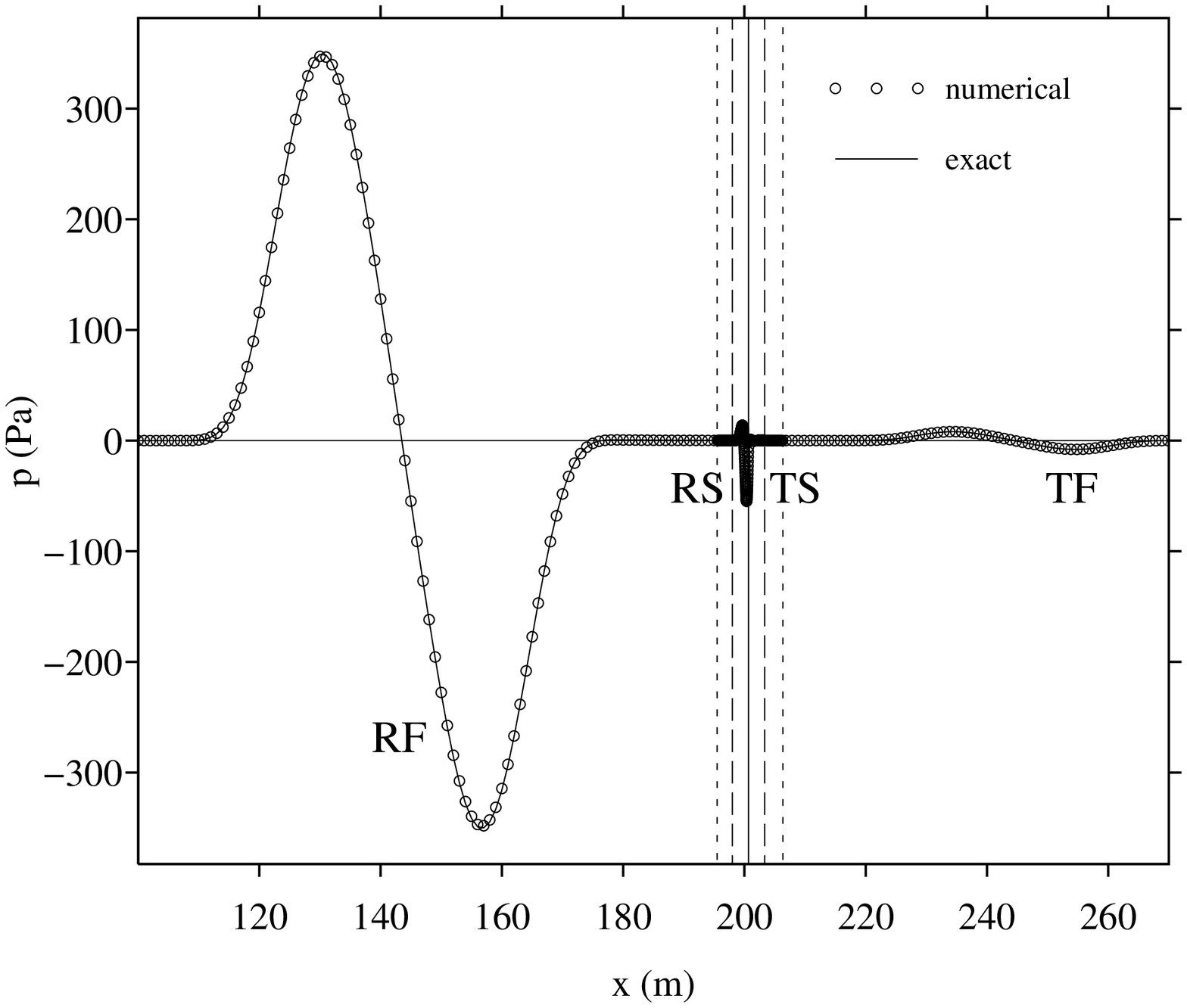}\\
(c) & (d)\\
\includegraphics[scale=0.32]{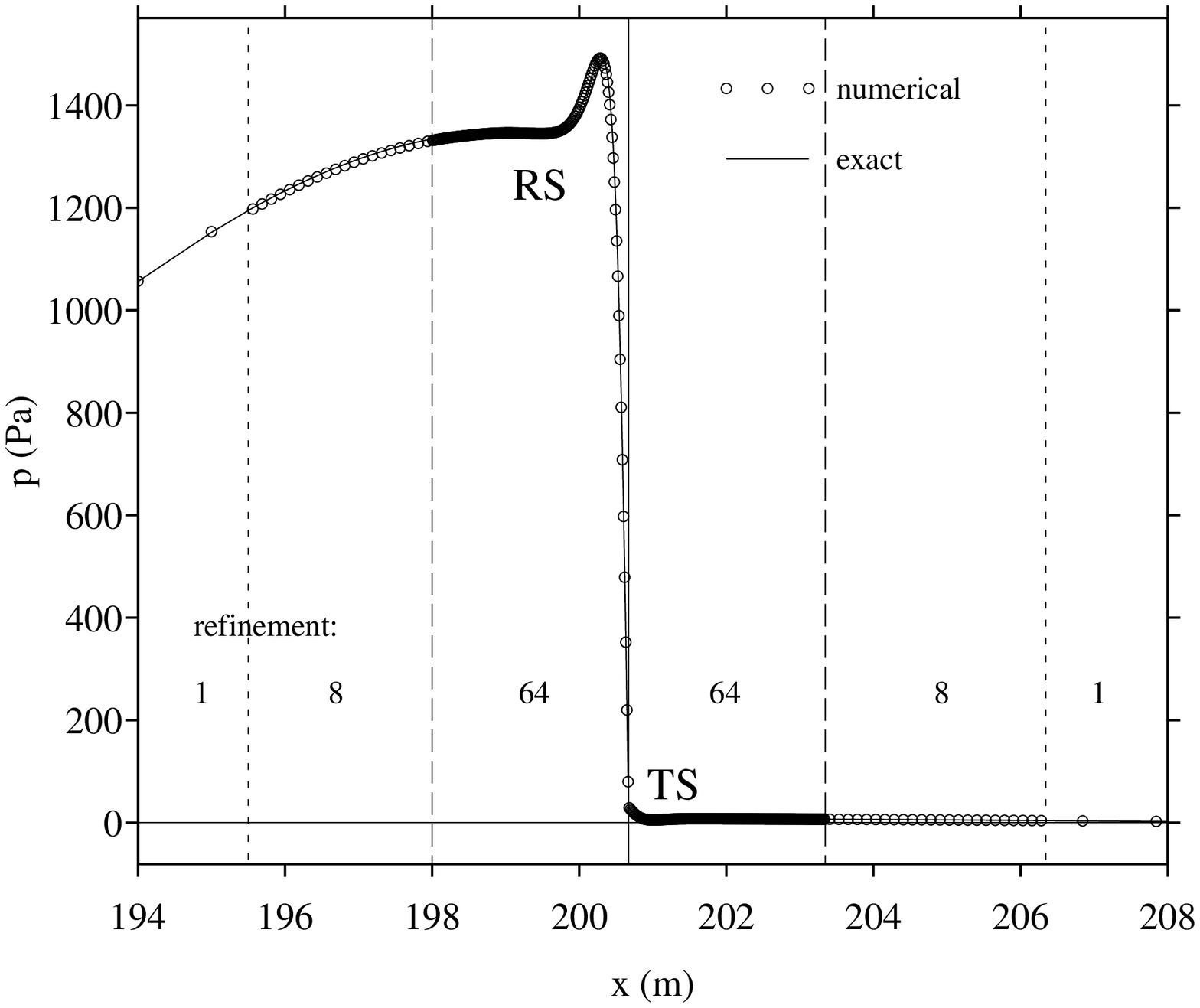}&
\includegraphics[scale=0.32]{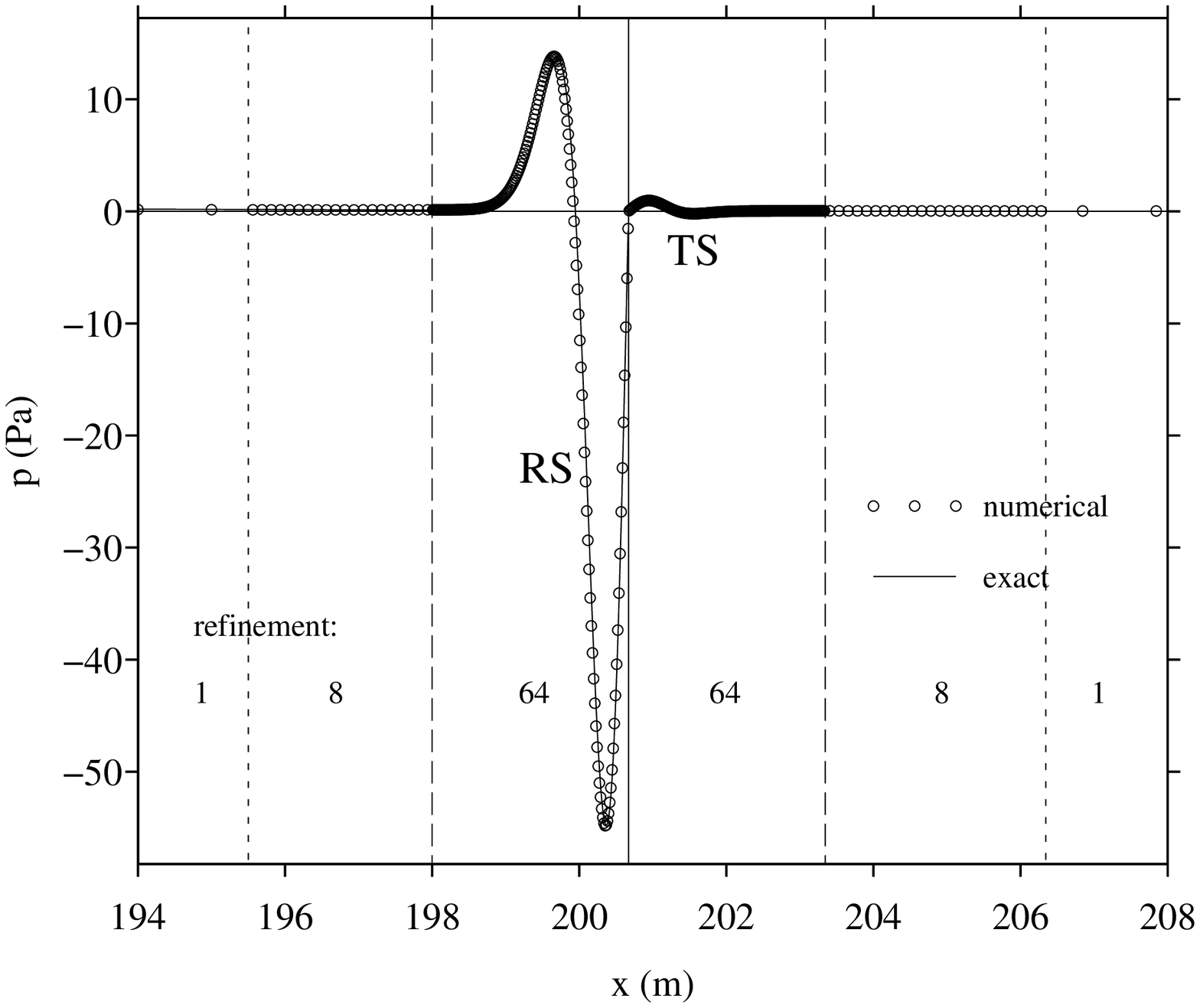}
\end{tabular}
\end{center}
\caption{Test 4: snapshot of $p$ at time $t_1$ (a-c) and $t_2$ (b-d). Zoom around $\alpha$ (c-d).} 
\label{Test4}
\end{figure}

Test 4 is performed on ($\Omega_0,\Omega_1$), with $1/\kappa_s=0$. Two successive mesh refinements with factors 8 and 64 are used around $\alpha$. Snapshots of $p$ at $t_1$ and $t_2>t_1$ are shown in figure \ref{Test4}. During the interaction of the incident fast wave with the interface (a-c), since the slow waves have a greater amplitude than that of fast waves, they play a crucial role in the balance of momentum and mass: a poor assessment of the slow waves would invalidate that of the other ones. At $t_2$ (b-d), the slow waves are greatly attenuated and remain localized near $\alpha$.

\begin{figure}[htbp]
\begin{center}
\begin{tabular}{cc}
(a) & (b)\\
\includegraphics[scale=0.32]{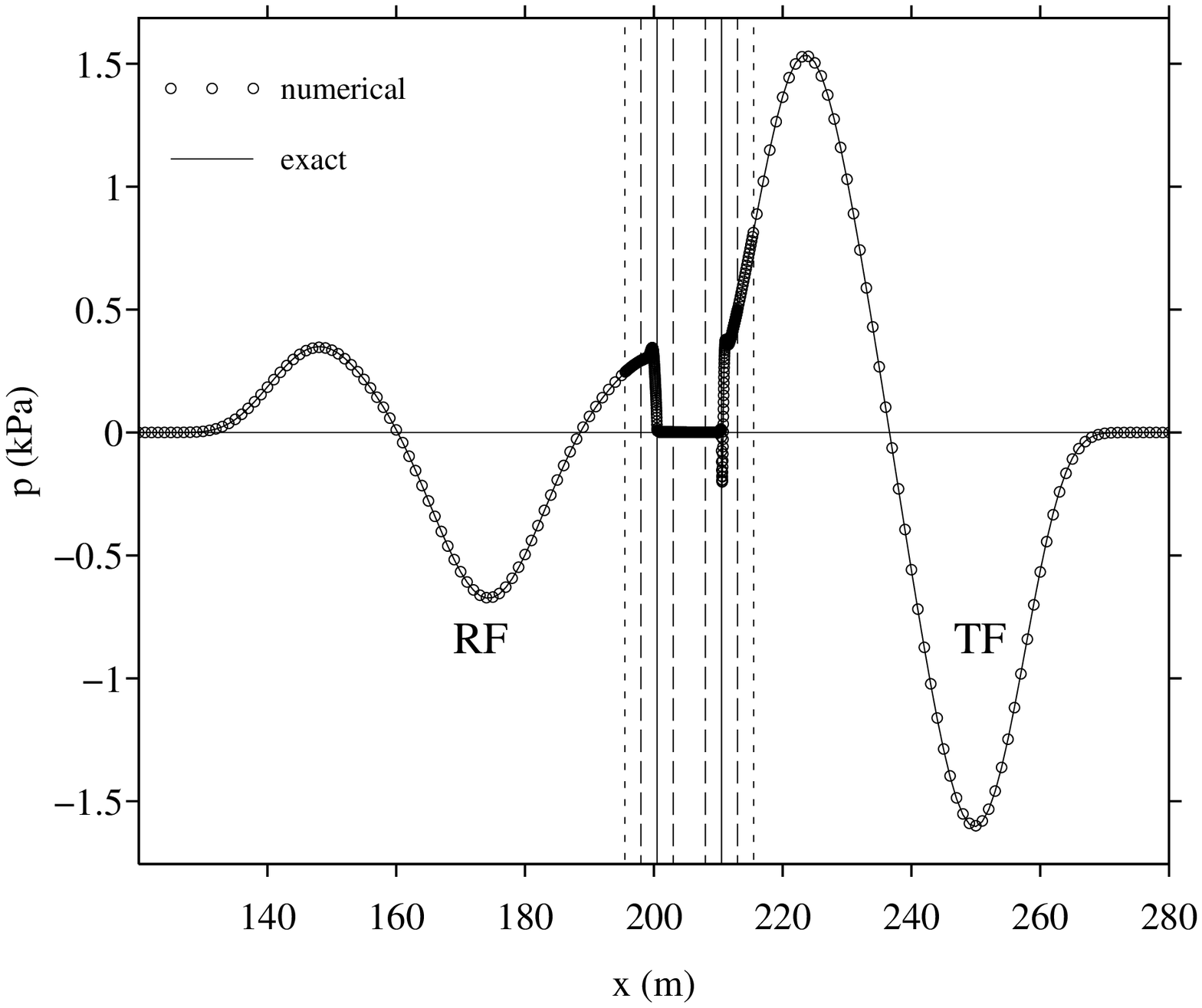}&
\includegraphics[scale=0.32]{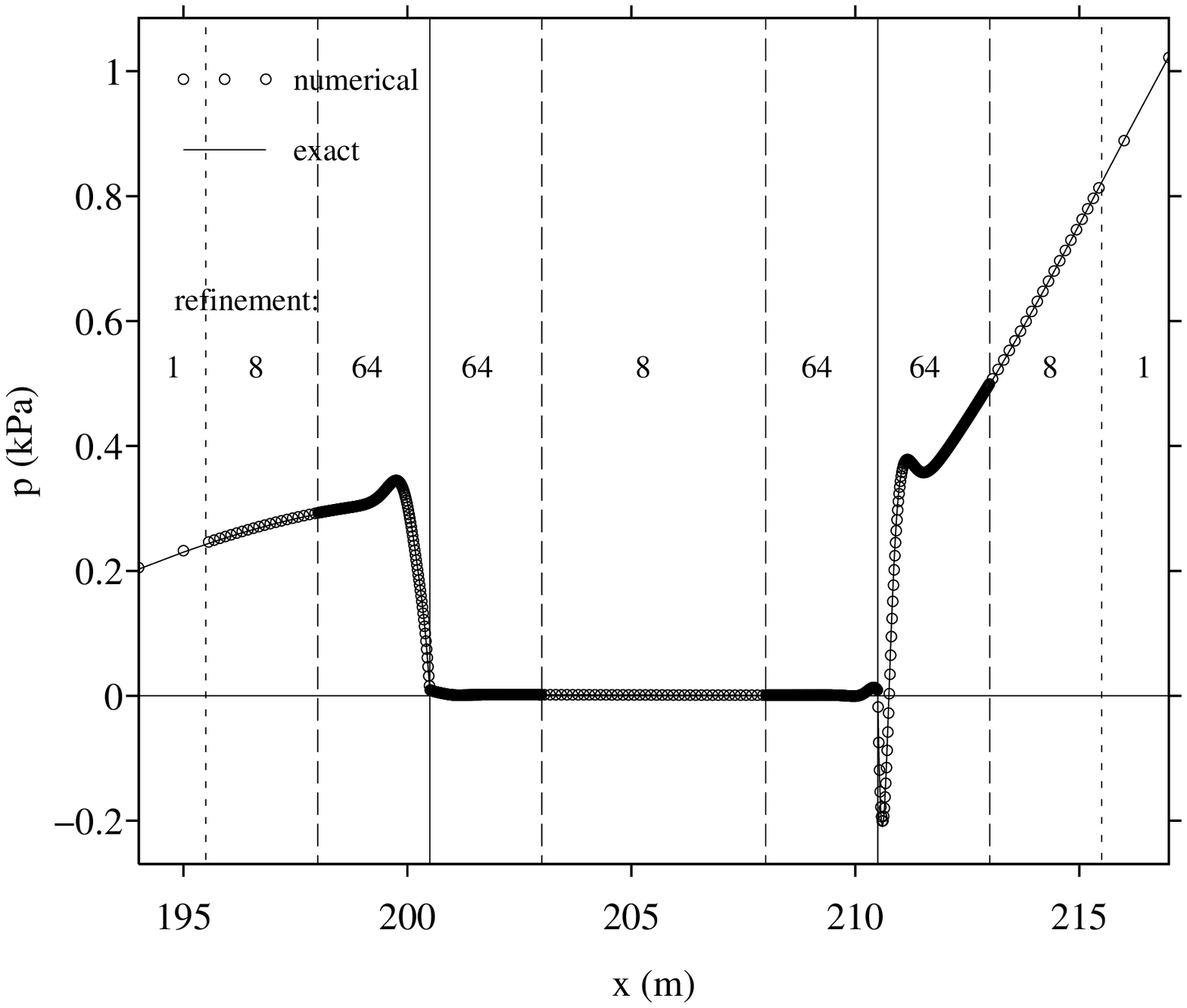}
\end{tabular}
\end{center}
\caption{Test 5: snapshot of $p$ after crossing the two interfaces (a), zoom (b).} 
\label{Test5}
\end{figure}

Lastly, two examples of wave propagation across multiple interfaces are investigated. The media $\Omega_0$ and $\Omega_1$ are repeated alternatively. Test 5 treats the case of two interfaces, with a known analytical solution (figure \ref{Test5}). As in test 4, one observes the small-scale evolution of the static slow waves generated by the interfaces.

\begin{figure}[htbp]
\begin{center}
\begin{tabular}{cc}
(a) & (b)\\
\includegraphics[scale=0.32]{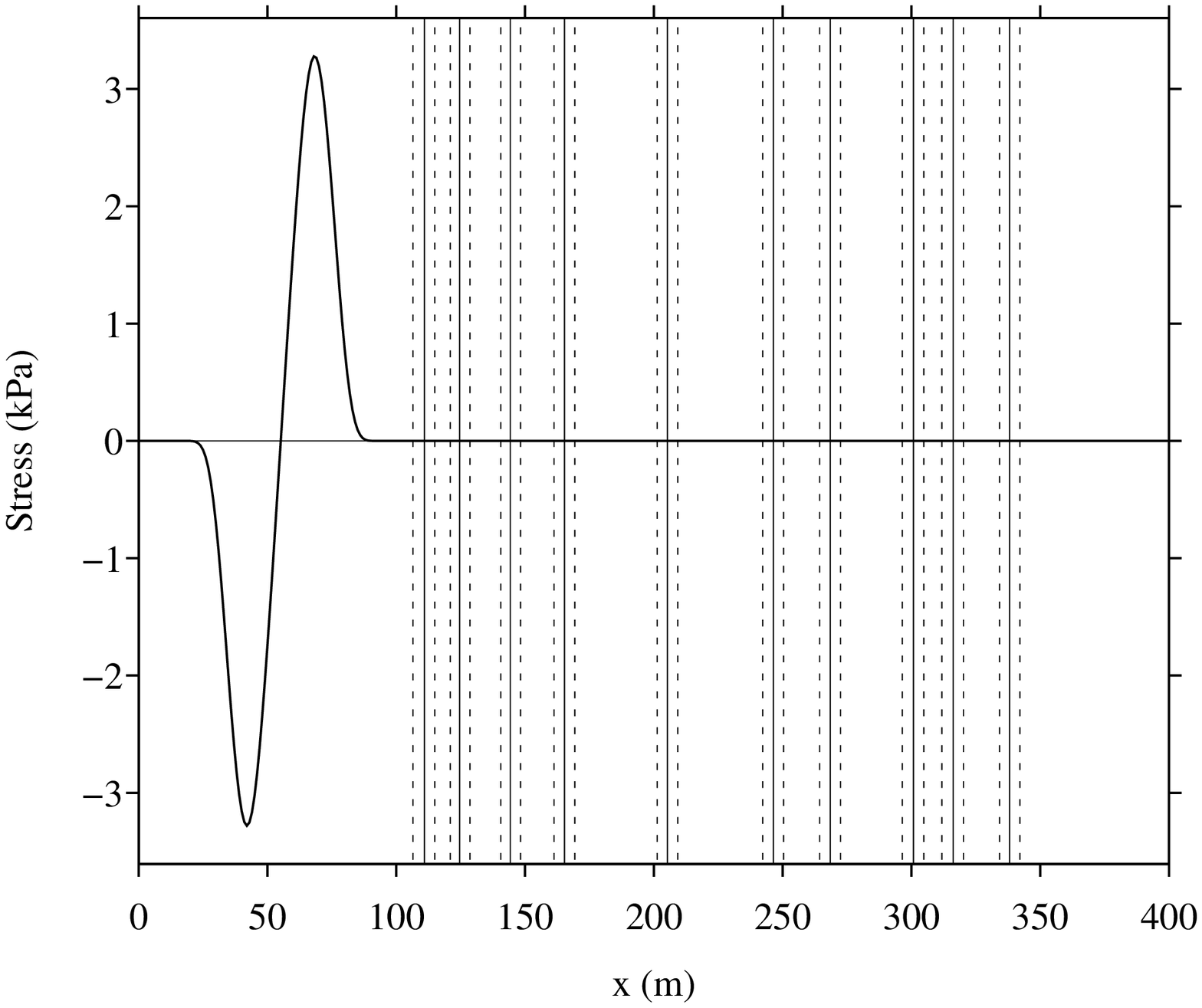}&
\includegraphics[scale=0.32]{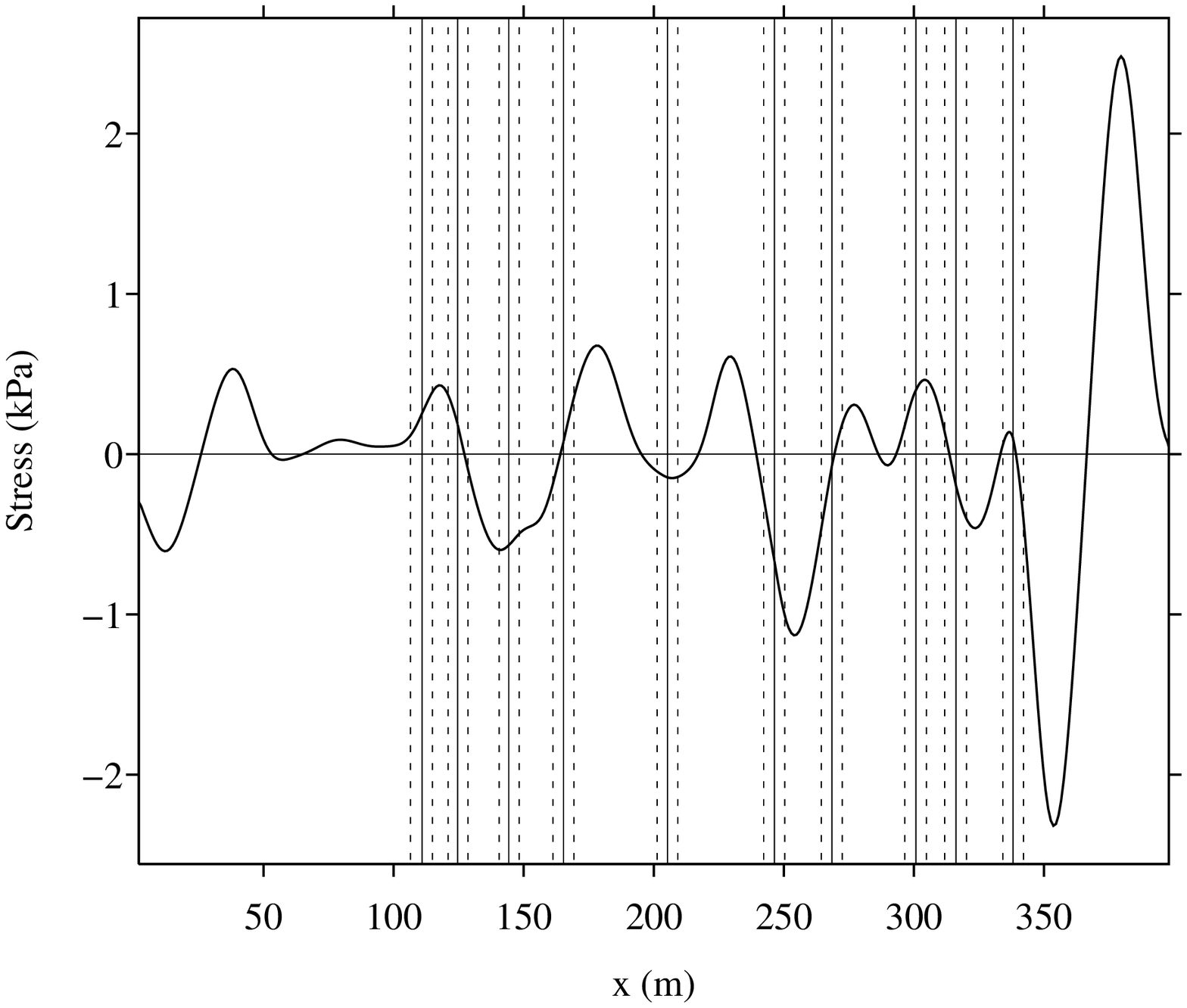}
\end{tabular}
\end{center}
\caption{Test 6: $\sigma$ at the initial instant (a) and after crossing the interfaces (b).} 
\label{Test6}
\end{figure}

Test 6 deals with a case of 10 interfaces with randomly distributed position. Performing such a simulation has physical applications, even in 1D \cite{LENOACH99}. Figure \ref{Test6} shows $\sigma$ at the initial instant (a) and after the wave has crossed the whole set of interfaces (b). For the sake of clarity, only grid refinements positions from 1 to 8 are shown.

\section{Conclusion}\label{SecConc}

Numerical modeling of 1-D transient Biot's model was addressed here for waves whose frequency content lie in the low-frequency range. Three numerical tools were combined to obtain a method describing accurately the wave propagation: a fourth-order scheme with time-splitting; a space-time mesh refinement; and an immersed interface method. This method is required to account for the properties of the slow wave. Further research is suggested:
\begin{itemize}
\item studying the high-frequency range \cite{BIOT56-B}, where $\kappa/\eta$ is proportional to $f^{1/2}$. Fractional derivatives are therefore involved in the time-domain \cite{MATIGNON07,HANYGA05};
\item accounting for dissipative effects in the solid skeleton;
\item coupling with nonlinear boundary conditions, to model seismic rupture;
\item extending the method to two-dimensional configurations. The validity of each of the numerical tools has already been established in 2D.
\end{itemize}

\end{document}